\documentclass[conference]{IEEEtran}
\usepackage[utf8]{inputenc}
\usepackage[noadjust]{cite}

\usepackage{graphicx}
\usepackage{amsmath}
\usepackage{amssymb}
\usepackage{mathtools}

\usepackage{caption}
\usepackage{subcaption}
\usepackage{pgfplotstable}
\usepackage{filecontents}
\usepackage{tikz, circuitikz}
\usepackage{pgfplots}
\usepackage{nicefrac}

\usepackage{textcomp}
\usepackage{hyperref}
\usepackage{lipsum}

\usepgfplotslibrary{units}
\usetikzlibrary{arrows}

\definecolor{Set1-7-1}{RGB}{228,26,28}
\definecolor{Set1-7-2}{RGB}{55,126,184}
\definecolor{Set1-7-3}{RGB}{77,175,74}
\definecolor{Set1-7-4}{RGB}{152,78,163}
\definecolor{Set1-7-5}{RGB}{255,127,0}
\definecolor{Set1-7-6}{RGB}{166,86,40}
\definecolor{Set1-7-7}{RGB}{0,0,0}

\definecolor{CGray}{RGB}{205,205,205}
\definecolor{CBlue}{RGB}{0,112,192}
\definecolor{COrange}{RGB}{228,108,10}
\definecolor{CGreen}{RGB}{133,181,57}
\definecolor{CRed}{RGB}{192,0,0}
\definecolor{CPurple}{HTML}{7e2f8e}
\definecolor{CYellow}{HTML}{edb120}

\DeclareMathOperator*{\argmax}{arg\,max}

\DeclareMathOperator*{\angleF}{angle}
\DeclareMathOperator*{\real}{Re}

\newcommand\copyrighttext{%
  \footnotesize \textcopyright 20XX IEEE. Personal use of this material is permitted.
  Permission from IEEE must be obtained for all other uses, in any current or future
  media, including reprinting/republishing this material for advertising or promotional
  purposes, creating new collective works, for resale or redistribution to servers or
  lists, or reuse of any copyrighted component of this work in other works.}
\newcommand\copyrightnotice{%
\begin{tikzpicture}[remember picture, overlay]
\node[anchor=south,yshift=10pt] at (current page.south) {\fbox{\parbox{\dimexpr\textwidth-\fboxsep-\fboxrule\relax}{\copyrighttext}}};
\end{tikzpicture}%
}

\begin{document}
\bstctlcite{IEEEexample:BSTcontrol}

\author{Mathieu Xhonneux, Jérôme Louveaux, and David Bol\\
    ICTEAM, UCLouvain, Belgium
}

\title{Implementing a LoRa Software-Defined Radio\\ on a General-Purpose ULP Microcontroller}
\maketitle
\copyrightnotice

\begin{abstract} 

Emerging Internet-of-Things sensing applications rely on ultra low-power (ULP) microcontroller units (MCUs) that wirelessly transmit data to the cloud.
Typical MCUs nowadays consist of generic blocks, except for the protocol-specific radios implemented in hardware.
Hardware radios however slow down the evolution of wireless protocols due to retrocompatiblity concerns.
In this work, we explore a software-defined radio architecture by demonstrating a LoRa transceiver running on custom ULP MCU codenamed SleepRider with an ARM Cortex-M4 CPU.
In SleepRider MCU, we offload the generic baseband operations (e.g., low-pass filtering) to a reconfigurable digital front-end block and use the Cortex-M4 CPU to perform
the protocol-specific computations. Our software implementation of the LoRa physical layer only uses the native SIMD instructions of the Cortex-M4 to
achieve real-time transmission and reception of LoRa packets.
SleepRider MCU has been fabricated in a 28nm FDSOI CMOS technology and is used in a testbed to experimentally validate the software implementation.
Experimental results show that the proposed software-defined radio requires only a CPU frequency of 20 MHz to correctly receive a LoRa packet,
with an ultra-low power consumption of 0.42~mW on average.
\end{abstract} 

\section{Introduction}
\label{sec:intro}

The last years have seen the rise of Internet-of-Things (IoT) smart sensing applications, in which low-power sensors running on a microcontroller unit (MCU) periodically send small information packets
to the cloud using a low-power wide-area network (LPWAN).
Yet, the foreseen massive deployment of IoT applications and its associated ecological footprint may prevent the ICT sector to become truly sustainable, especially when considering that Moore's law is slowing down~\cite{bol2021moore}.
Although IoT sensors have a very low energy consumption, they still own a non-negligible ecological footprint, mainly due to the embodied energy used during their fabrication~\cite{pirson2021}.

Nowadays, IoT MCUs usually consist of generic blocks (processor, memories, ADC, ...), at the exception of the radios.
The physical layer (PHY) of LPWAN radios are commonly implemented in hardware and are hence
tied to one or several specific protocols. This historical paradigm however features several issues.
First, IoT protocols cannot be updated without disposing of the already deployed sensor networks and deploying new devices.
Similarly, the re-use of sensors for another application with different communication requirements (e.g., greater range or higher throughput) is impossible.
Even more, we argue that the current paradigm hinders the improvement of LPWAN standards due to the need of maintaining retrocompatiblity with existing sensors.
The most popular LPWAN technologies deployed as of today are LoRa and SigFox~\cite{haxhibeqiri2018survey}.
Yet, both the LoRa and the SigFox specifications have barely evolved since their initial rollout in the early 2010s,
despite the significant research conducted on LPWANs in the last decade.

IoT sensors with software-defined-radios (SDRs) have the potential to overcome these issues~\cite{chen2016low,amor2019software}.
In an SDR, most of the digital baseband computations are performed in software by a processor (CPU).
Such an architecture enables on-the-fly updates of the communication protocol, and therefore allows a radio to switch
from one standard to another when the requirements of the application evolve, or simply to follow the latest revisions of a protocol.
The flexibility offered by SDRs could therefore help to both extend the lifetime of a sensor and foster innovation at the physical layer of LPWANs.

\begin{figure}[t]
    \centering
    \includegraphics[width=0.95\linewidth]{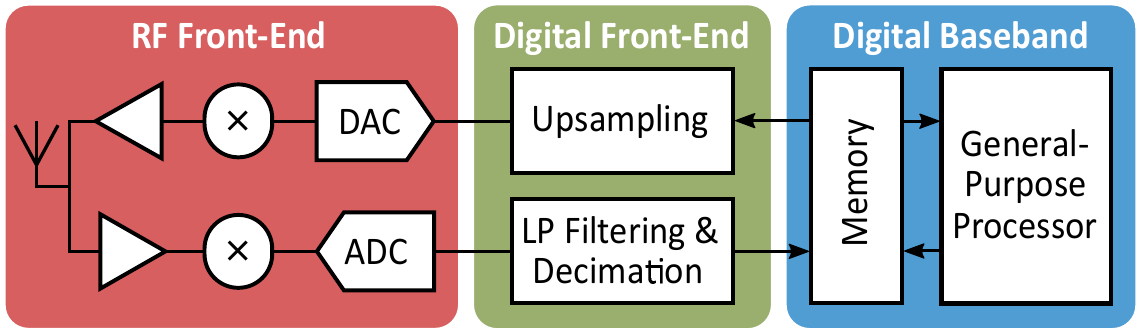}
    \caption{Architecture of an IoT software-defined radio with a common sub-GHz RF front-end, a reconfigurable digital front-end (DFE)
    and a general-purpose processor performing the protocol-specific digital baseband (DBB) computations.}
    \label{fig:global-archi}
\end{figure}

Nevertheless, the main challenge of IoT SDRs is to transmit and receive packets at the specified signaling rate while preserving a low-power consumption.
Two different approaches have been identified to tackle this issue.
In~\cite{chen2016low}, the baseband computations are relegated to a custom baseband processor with an instruction set specifically designed for IoT protocols.
In~\cite{amor2019software},
the authors observe that LPWANs share very similar characteristics (i.e., the use of small bandwidths in sub-GHz bands),
and therefore propose an architecture (shown in Fig.~\ref{fig:global-archi}) with a single common RF front-end, a reconfigurable digital front-end (DFE) that executes the generic
baseband operations (e.g., low-pass filtering), and a general-purpose processor that performs the protocol-specific computations.

In this paper, we follow the latter approach, which is the most flexible in our opinion, to implement a functional LoRa-compliant SDR
on a general-purpose ultra-low-power (ULP) MCU.
Whereas~\cite{amor2019software} evaluate in simulation the feasibility of demodulating generic GFSK signals on several processors,
we present a real-time LoRa PHY implementation on a custom ULP MCU prototyped in 28nm FDSOI CMOS featuring a DFE and an ARM Cortex-M4 CPU.

The remainder of this paper is organized as follows. We first briefly describe the LoRa protocol in Section~\ref{sec:lora}.
We then explain the architecture of our SDR-capable MCU in Section~\ref{sec:hw},
and we describe our software implementation of the LoRa PHY in Section~\ref{sec:sw}.
Finally, the performance of our SDR is experimentally evaluated in Section~\ref{sec:perfs}.

\section{The LoRa Physical Layer}
\label{sec:lora}

LoRa is an IoT physical layer (PHY) that operates within the unlicensed sub-GHz ISM frequency bands~\cite{haxhibeqiri2018survey}.
In this section, we present the different blocks of the LoRa PHY and describe the computations carried out by a typical LoRa transceiver.

\subsection{Modulation and Demodulation}

LoRa uses a chirp spread-spectrum modulation, for which each symbol corresponds to a chirp, i.e., a complex phasor whose instantaneous frequency increases linearly with time.
The duration $T_S$ of a symbol is $T_S = \frac{2^{\mathrm{SF}}}{B}$, where $\mathrm{SF}$ is the \textit{spreading factor} and $B$ is the bandwidth of the chirp.
Selecting a large spreading factor increases the symbol period, which in turn decreases the data throughput but enhances the communication range~\cite{haxhibeqiri2018survey}.
LoRa transceivers typically sample chirps at the Nyquist frequency $f_S = B$, implying that each symbol contains $N = 2^{\mathrm{SF}}$ samples.
The range of valid spreading goes from 7 to 12~\cite{haxhibeqiri2018survey}.

LoRa symbols are modulated by selecting the initial instantaneous frequency of the chirp, with $N$ possible different initial frequencies. The complex baseband-equivalent Nyquist-rate representation of
a symbol $x_s[n]$ modulated with a symbol $s$ between $0$ and $N-1$ is given by
\begin{equation} \label{eq:symbol-discrete}
    x_s[n] = e^{j 2 \pi \left[\frac{n^2}{2 N} + \left( \frac{s}{N} - \frac{1}{2} \right) n \right]},
\end{equation}
where $n \in \{0, \dots, N-1\}$ is the sample index~\cite{ghanaatian2019lora}.
Thanks to time/frequency equivalence properties of the chirps, a LoRa symbol $s$ can be efficiently modulated
by performing a cyclic shift of $s$ samples on the base waveform $x_0[n]$~\cite{elshabrawy2019different}:
\begin{equation} \label{eq:symbol-mod}
    x_s[n] = x_0[(n-s) \bmod N].
\end{equation} 

A LoRa receiver implements the following steps to demodulate a symbol.
Let $y_s[n] = x_s[n] + w[n]$ be the received signal when the receiver is perfectly synchronized in time and frequency,
where $w[n]$ is additive white Gaussian noise (AWGN).
The receiver processes windows of $N$ samples, with each window containing one symbol.
For every window, the sampled signal $y_s[n]$ is first multiplied point-wise with $\overline{x}_0[n]$, the complex conjugate of the base waveform.
Multiplying the received chirp by $\overline{x}_0[n]$ is called \textit{dechirping}, 
as it removes the squared phase component from $y_s[n]$ but leaves the frequency term that depends on $s$ which carries the modulated information.
The dechirped signal $\tilde{y}_s[n]$ contains a single-tone term of frequency $\frac{s}{N}$ and AWGN such as
\begin{equation}
    \label{eq:dechirping}
    \tilde{y}_s[n] = y_s[n] \cdot \overline{x}_0[n] = e^{j 2 \pi \frac{sn}{N}} + \tilde{w}[n],
\end{equation}
where $\tilde{w}[n] = \overline{x}_0[n] \cdot w[n]$. 
The maximum-likelihood detector computes the $N$-point discrete Fourier transform (DFT) of $\tilde{y}_s[n]$~\cite{ghanaatian2019lora}.
Let $Y_k$ be the DFT of the dechirped signal $\tilde{y}_s[n]$.
In a noiseless scenario with perfect synchronization, $Y_k$ contains a single peak of height $N$ at the index $k = s$. 
To avoid a tracking of the phase, LoRa receivers commonly implement the non-coherent maximum-likelihood rule,
which selects the DFT bin of greatest energy, i.e., $\widehat{s} = \argmax_k \left| Y_k \right|$~\cite{ghanaatian2019lora}.

\subsection{Synchronization and Preamble Structure}

To demodulate a LoRa packet, the receiver first needs to synchronize with the transmitter. The synchronization of a LoRa receiver requires the correction
of a carrier frequency offset (CFO) $\Delta f_c$ and a sampling time offset (STO) $\tau$~\cite{bernier2020low}.
A receiver that is not synchronized in time retrieves windows of $N$ samples containing two consecutive partial chirps instead of a single entire chirp. The
STO $\tau$ represents the sample-level time offset between the beginning of the window and the first sample of the second chirp in the window.
The CFO $\Delta f_c$ represents the difference of carrier frequency between the transmitter and the receiver.
Due to specificities of the chirp spread-spectrum modulation, the CFO and the STO can be decomposed into integer and fractional components:
\begin{equation}
    \label{eq:deltafc-tau}
    \Delta f_c = \frac{B}{N} \left( L_{\text{CFO}} + \lambda_{\text{CFO}} \right), \tau = L_{\text{STO}} + \lambda_{\text{STO}},
\end{equation}
where $L_{\text{CFO}}$, $L_{\text{STO}}$ are integer numbers and $\lambda_{\text{CFO}}, \lambda_{\text{STO}}$ are fractional values in the range $]-0.5, 0.5]$~\cite{bernier2020low,xhonneux2019low}.
Integer offsets shift the peak of a symbol after the DFT, initially located at the index $k = s$, to the bin $k = \left( s + L_{\text{STO}} + L_{\text{CFO}} \right) \bmod N$.
On the other hand, fractional offsets scatter the energy of the symbol previously contained in a single bin over adjacent frequency bins.
This scattering reduces the probability of correctly demodulating a symbol in the presence of AWGN~\cite{xhonneux2019low}.

To facilitate the synchronization, all LoRa packets start with a preamble, whose structure is illustrated in Fig.~\ref{fig:tx-rx-chains}~\cite{tapparel2020open}.
The preamble contains eight repetitions of an \textit{upchirp} (the base waveform $x_0[n]$), followed by two
network identifier symbols and $2.25$ repetitions of a \textit{downchirp}.
Downchirps are chirps whose instantaneous frequency decreases with time, i.e., the complex conjugate $\overline{x}_0[n]$ of the base waveform
and can be demodulated similarly to an upchirp by dechirping the sampled signal with its complex conjugate, i.e., an upchirp $x_0[n]$.
The joint usage of upchirps and downchirps in the preamble is required to separetely estimate the integer offsets~\cite{bernier2020low}.
The integer CFO can be estimated by summing the demodulated values $s_{\text{up}}$ and $s_{\text{down}}$ of an upchirp and a downchirp, respectively,
and dividing their sum by two. The integer STO is then obtained by subtracting the estimate of $L_{\text{CFO}}$ from $s_{\text{up}}$.

\subsection{Complete Tx and Rx Chains}

\begin{figure}[t]
    \centering
    \includegraphics[width=\linewidth]{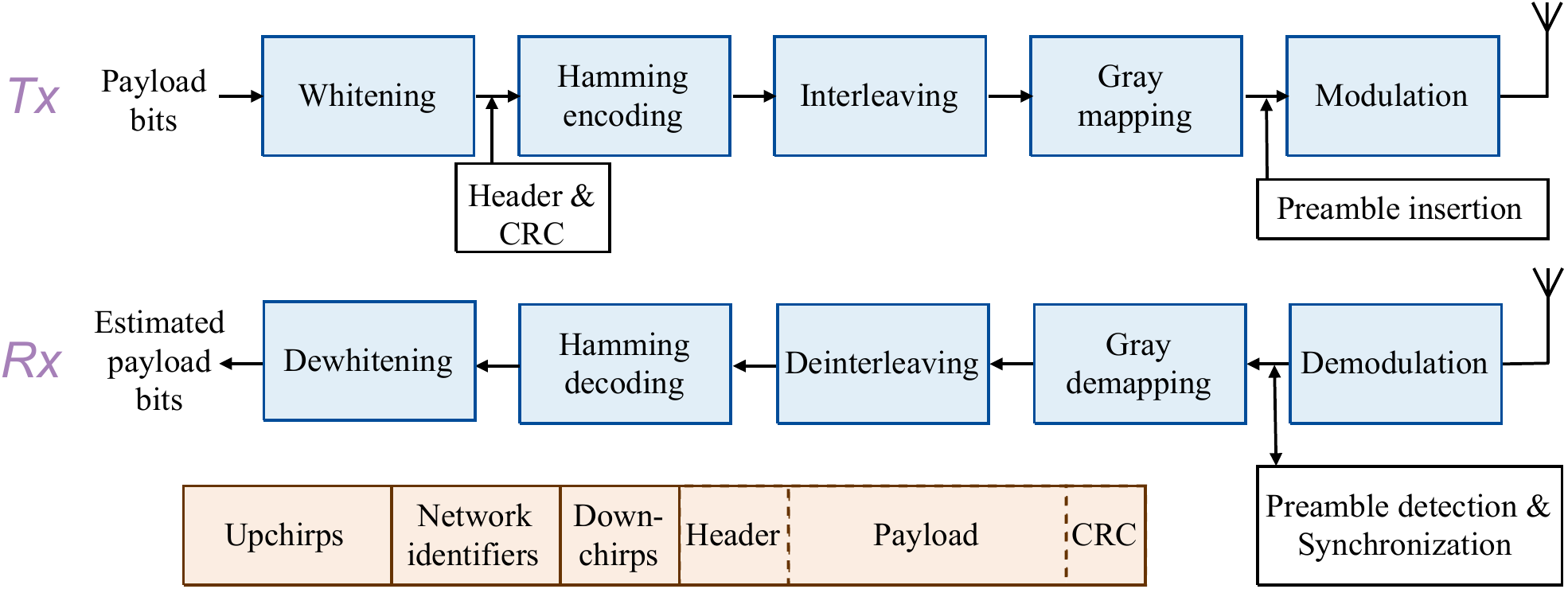}
    \caption{Transmit and receive chains of a LoRa transceiver (in blue), and structure of LoRa packets (in brown).}
    \label{fig:tx-rx-chains}
\end{figure}

Beside the modulation and demodulation stages, LoRa transceivers also perform some additional processing, as shown in Fig.~\ref{fig:tx-rx-chains}.
At the transmitter, the input payload bits are first whitened with a predefined pseudo-random sequence~\cite{tapparel2020open}.
A header is also inserted before the whitened payload. This header indicates to the receiver the length of the payload, the chosen coding rate and the presence of an optional 16-bit cyclic redundancy check (CRC)
at the end of the packet. At this point, all bits in the packet are coded using a $(k, n)$ Hamming code with $k = 4$ and $n \in \{6,7,8\}$~\cite{tapparel2020open}.
The coded bits are then interleaved using a diagonal interleaver and, finally, the interleaved bits are mapped to symbols with a Gray code.
The Rx chain starts when a preamble is detected. The receiver then synchronizes to the transmitter by estimating and correcting its CFO and STO.
The received symbols are subsequently demodulated and mapped back to bits. After the reception of the entire packet, the receiver
performs the deinterleaving, Hamming decoding and dewhitening. 

\section{Microcontroller Architecture for IoT Software-Defined Radios}
\label{sec:hw}

\begin{figure}[t]
    \centering
    \includegraphics[width=0.95\linewidth]{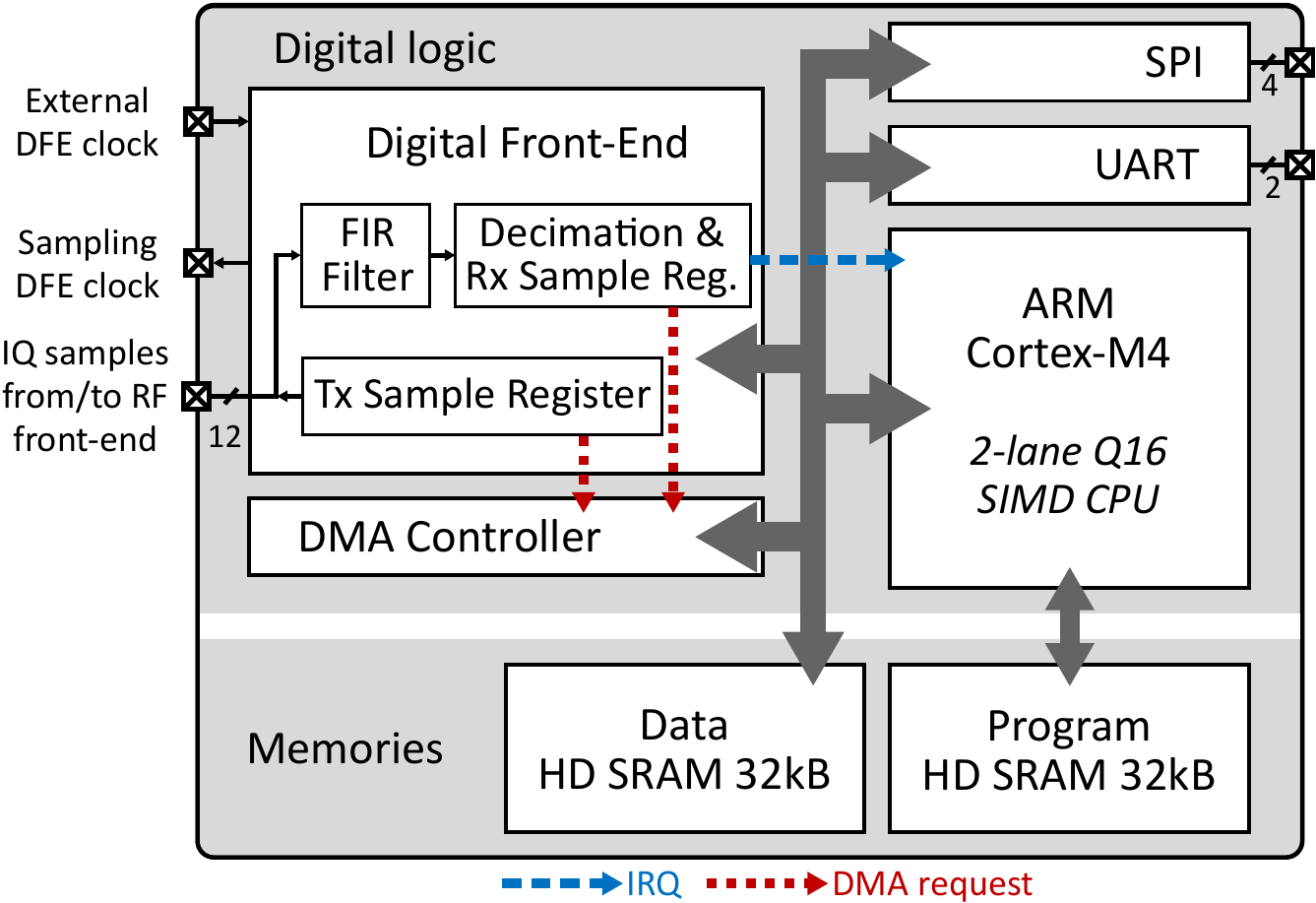}
    \caption{Architecture of our SDR-capable microcontroller.}
    \label{fig:archi}
\end{figure}

In this section, we present the architecture of our ULP MCU codenamed SleepRider, illustrated in Fig.~\ref{fig:archi}, which has been designed to run SDR implementations of IoT protocols.
The MCU embeds an ARM Cortex-M4 CPU, two high-density 32kB SRAM memories, a Direct Memory Access (DMA) controller, a digital front-end (DFE)
and several I/O peripherals.
The Cortex-M4 follows a Harvard architecture, i.e., one SRAM acts as a program memory (PMEM) and stores the assembly instructions, whereas the
other one is used by the CPU as a data memory (DMEM). Except for the PMEM, the CPU and all other peripherals share a single memory bus to exchange data.
SleepRider MCU was designed for ULP in 28nm FDSOI with various techniques including ultra-low supply voltage (0.4V) for its logic.

The Cortex-M4 CPU is a 32-bit processor designed for low-power embedded applications and is a popular choice for commercial IoT MCUs.
It features a three-stage pipeline and two single-instruction multiple-data (SIMD) lanes.
SIMD instructions notably allow the CPU to efficiently execute DSP computations on pairs of Q16 fixed-point numbers.
For instance, the $\texttt{SMUAD}$ instruction performs two multiplications on two pairs of Q16 numbers and returns their sum in a single cycle.
Such instructions are particularly useful to speed up complex baseband operations (e.g., FFTs) when complex samples are coded as
pairs of Q16 numbers.

Our MCU architecture follows the one from~\cite{amor2019software}, with a DFE that interfaces the RF transceiver with the DMEM.
The DFE is a custom digital block that contains separate datapaths and control logic for the Tx and Rx chains.
In Rx, the block retrieves digital samples from the RF chain at an oversampled frequency $R f_S$.
The rate at which the DFE retrieves samples, either in Tx or Rx, is hereafter called the DFE frequency.
The oversampled samples are first low-pass filtered with a FIR filter and then decimated to the baseband frequency $f_S$.
The oversampling by a factor $R$ at the input allows to achieve fine-grain time synchronization. When demodulating samples,
the CPU selects the decimation index among the $R$ parallel streams that minimize the sampling time offset.
In Tx, the DFE directly transfers samples to the RF front-end without performing upsampling, i.e.,
the DFE frequency is set to the baseband frequency at which the CPU modulates symbols.

In either transmission or reception, the digital samples retrieved from or sent to the RF chain are complex baseband samples of 24-bit, i.e.,
the real and imaginary parts are both coded on 12 bits.
Since the CPU contains two 16-bit SIMD lanes, these 12-bit words are extended to (resp. truncated from) 16 bits
before entering (resp. leaving) the DMEM.

To alleviate the CPU load, the DMA controller is responsible for transferring the 32-bit complex baseband samples
between the DMEM and the DFE. The memory transfers are organized as follows.
The CPU first defines windows of $L$ complex samples in the DMEM and indicates the length of these windows to the DFE.
In Tx, when the CPU has filled a window with modulated samples, it configures the DMA to transfer the $L$ 32-bit words to the DFE.
The rate of the memory transfers is determined by the DFE frequency, i.e., the DFE processes one complex sample at a time and only
requests a new sample to the DMA controller when it has transmitted the previous one to the RF front-end.
Similarly, in Rx, the CPU first configures the DMA controller for $L$ transfers, and the DFE then issues for each new decimated sample a 32-bit transfer request to the controller.
When $L$ samples have been transferred, the DFE raises an interrupt request (IRQ) to the CPU,
indicating that the window has been transmitted (in Tx) or is ready to be demodulated (in Rx).

Since the proposed design aims to enable the implementation of multiple IoT protocols
with different requirements (e.g., bandwidth, signaling rates), the DFE has been designed to be as flexible as possible.
First, the DFE frequency is generated by an external clock, i.e., any frequency can be attained with a dedicated clock generator.
Frequency dividers in the DFE further allow the programmer to divide the external clock to obtain different sampling rates.
The decimation rate at the output of the FIR filter can also be configured to obtain a specific baseband sampling frequency.
Finally, the weights of the FIR filter are configurable and can be defined following the waveform specifications.

\section{Software Implementation of the LoRa PHY}
\label{sec:sw}

In this section, we describe our software implementation of the LoRa PHY for the MCU architecture presented in Section~\ref{sec:hw}.
Our implementation is coded in C and relies on the ARM CMSIS DSP library, which provides many complex DSP routines optimized for the Cortex-M4.
We first explain the Tx chain and then describe the Rx one. The latter is notably more complicated because of the synchronization stage.

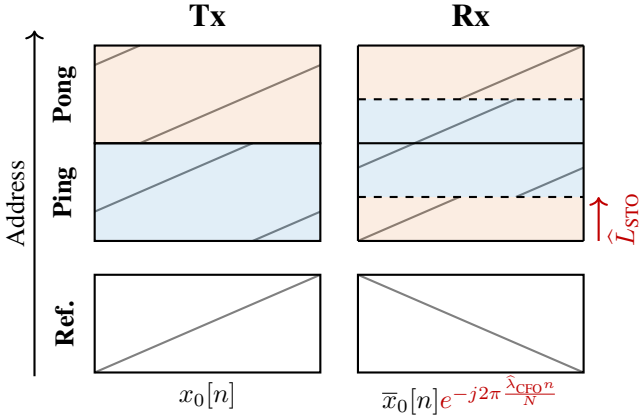
\begin{figure}[t]
    \centering
    \pgfmathsetmacro{\bH}{1.3}
\pgfmathsetmacro{\bW}{3}
\pgfmathsetmacro{\sA}{atan(\bH / \bW)}

\pgfmathsetmacro{\oW}{0.5}
\pgfmathsetmacro{\oH}{0.45}

\pgfmathsetmacro{\SMA}{0.3}
\pgfmathsetmacro{\SMB}{0.8}

\pgfmathsetmacro{\SDA}{0.3}
\pgfmathsetmacro{\SDB}{0.55}

\pgfmathsetmacro{\tau}{0.45}

\begin{tikzpicture}

    \draw [->,thick] (-0.8, -0.05) -- (-0.8, \oH + 3*\bH + 0.2) node [midway, above, sloped] (Addr) {Address};
    \draw [->,thick,color=CRed] (\oW + 2*\bW + 0.2, \oH + \bH) -- (\oW + 2*\bW + 0.2, \oH + \bH + \tau*\bH) node [midway, below, sloped] (Tau) {\textcolor{CRed}{$\widehat{L}_{\text{STO}}$}};

    \node[text centered, font=\bfseries] at (0.5 * \bW, \oH + 3*\bH + 0.4) {\large Tx};
    \node[text centered, font=\bfseries] at (1.5 * \bW + \oW, \oH + 3*\bH + 0.4) {\large Rx};

    \node[text centered, font=\bfseries, rotate=90] at (-0.4, 0.5*\bH) {Ref.};
    \node[text centered, font=\bfseries, rotate=90] at (-0.4, \oH + 1.5*\bH) {Ping};
    \node[text centered, font=\bfseries, rotate=90] at (-0.4, \oH + 2.5*\bH) {Pong};

    \node[text centered] at (0.5 * \bW, -0.3) {$x_0[n]$};
    \node[text centered] at (1.5 * \bW + \oW, -0.3) {$\overline{x}_0[n] \textcolor{CRed}{e^{-j 2 \pi \frac{\widehat{\lambda}_{\text{CFO}} n}{N}}}$};

    \draw[gray, thick] (0,0) --  (\bW, \bH);
    \draw[thick] (0,0) rectangle (\bW, \bH);

    \draw[gray, thick] (0, \oH + \bH + \bH*\SMA) -- (\bW - \SMA*\bW, \oH + 2*\bH);
    \draw[gray, thick] (\bW - \SMA*\bW, \oH + \bH) -- (\bW, \oH + \bH + \bH*\SMA);
    \draw[thick, draw=black, fill=CBlue, fill opacity=0.12] (0, \oH + \bH) rectangle (\bW, \oH + 2*\bH);

    \draw[gray, thick] (0, \oH + 2*\bH + \bH*\SMB) -- (\bW - \SMB*\bW, \oH + 3*\bH);
    \draw[gray, thick] (\bW - \SMB*\bW, \oH + 2*\bH) -- (\bW, \oH + 2*\bH + \bH*\SMB);
    \draw[thick, draw=black, fill=COrange, fill opacity=0.12] (0, \oH + 2*\bH) rectangle (\bW, \oH + 3*\bH);

    \draw[gray, thick] (\oW + \bW, \bH) --  (\oW + 2*\bW, 0);
    \draw[thick] (\oW + \bW, 0) rectangle (\oW + 2*\bW, \bH);

    \draw[gray, thick] (\oW + \bW, \oH + \bH + \bH*\SDA + \tau*\bH) -- (\oW + 2*\bW - \SDA*\bW, \oH + 2*\bH + \tau*\bH);
    \draw[gray, thick] (\oW + 2*\bW - \SDA*\bW, \oH + \bH + \tau*\bH) -- (\oW + 2*\bW, \oH + \bH + \bH*\SDA + \tau*\bH);

    \draw[gray, thick] (\oW + \bW, \oH + 2*\bH + \bH*\SDB + \tau*\bH - 2*\bH) -- (\oW + 2*\bW - \SDB*\bW, \oH + 3*\bH + \tau*\bH - 2*\bH);
    \draw[gray, thick] (\oW + 2*\bW - \SDB*\bW, \oH + 2*\bH + \tau*\bH) -- (\oW + 3*\bW - \SDB*\bW - \tau*\bW, \oH + 3*\bH);
    \path[fill=CBlue, fill opacity=0.12] (\oW + \bW, \oH + \bH + \tau*\bH) rectangle (\oW + 2*\bW, \oH + 2*\bH + \tau*\bH);
    \path[fill=COrange, fill opacity=0.12] (\oW + \bW, \oH + \bH) rectangle (\oW + 2*\bW, \oH + \bH + \tau*\bH);
    \path[fill=COrange, fill opacity=0.12] (\oW + \bW, \oH + 2*\bH  + \tau*\bH) rectangle (\oW + 2*\bW, \oH + 3*\bH);

    \draw[thick, dashed] (\oW + \bW, \oH + \bH + \tau*\bH) -- (\oW + 2*\bW, \oH + \bH + \tau*\bH);
    \draw[thick, dashed] (\oW + \bW, \oH + 2*\bH + \tau*\bH) -- (\oW + 2*\bW, \oH + 2*\bH + \tau*\bH);
    \draw[thick] (\oW + \bW, \oH + 2*\bH) -- (\oW + 2*\bW, \oH + 2*\bH);
    \draw[thick] (\oW + \bW, \oH + \bH) -- (\oW + 2*\bW, \oH + \bH);
    \draw[thick] (\oW + \bW, \oH + 3*\bH) -- (\oW + 2*\bW, \oH + 3*\bH);
    \draw[thick] (\oW + \bW, \oH + \bH) -- (\oW + \bW, \oH + 3*\bH);
    \draw[thick] (\oW + 2*\bW, \oH + \bH) -- (\oW + 2*\bW, \oH + 3*\bH);

    
\end{tikzpicture}
    \caption{Organization of the buffers in the DMEM for transmission (Tx) and reception (Rx).
    The gray lines in the buffers show the instantaneous frequency of the modulated chirps.}
    \label{fig:sw-memory}
\end{figure}

\subsection{Tx Implementation}

In the Tx chain, all bit-level operations (whitening, Hamming coding, interleaving and Gray mapping) are straightforwardly implemented in C.
To modulate LoRa symbols, we use the property given in Eq. \ref{eq:symbol-mod} with a reference waveform $x_0[n]$ stored in a buffer in the DMEM.
The waveform $x_s[n]$ of a symbol $s$ can then be obtained by copying the sequence $x_0[n]$ with a cylic shift of $s$ samples.
To transmit symbols in real time, the software relies on two ping-pong buffers.
The CPU computes modulated samples and stores them in one buffer, while the DMA controller transfers the samples from the other buffer
to the DFE. After the transmission of an entire buffer, the DFE notifies the CPU with an IRQ and the roles of the ping-pong buffers are exchanged.
If the CPU fills one buffer before the transmission of the other buffer,
it goes to sleep and waits for an IRQ of the DFE.
The DFE frequency is set at the Nyquist frequency $f_S = B$ and the ping-pong buffers hence contain a single LoRa symbol of $N$ complex samples each.
Figure~\ref{fig:sw-memory} shows the organization of the DMEM, in which all three buffers are stored.
The reference waveform $x_0[n]$ is generated only once in the Q16 fixed-point format, when the Tx chain is initialized.

\subsection{Rx Implementation}

Regarding the Rx chain, our receiver notices the arrival of a new packet when it consecutively demodulates three
identical symbols $\hat{s}$ (or its adjacent values $\hat{s} \pm 1$), similarly to~\cite{tapparel2020open}. To perform a fast demodulation of the received samples,
the Rx chain also uses two adjacent ping-pong buffers and a buffer storing $\overline{x}_0[n]$, the conjugate of the base waveform, as shown in Fig.~\ref{fig:sw-memory}.
The DFE is configured to decimate the received signal at the Nyquist frequency $f_S = B$ after the low-pass filtering.
When a buffer is full, the DFE notifies the CPU with an IRQ and the processor starts its demodulation by dechirping the
sampled signal with the sequence $\overline{x}_0[n]$ stored in memory and computing the FFT of the dechirped symbol.
All demodulation computations are executed in Q16 fixed-point to benefit from the SIMD instructions of the CPU.
Similarly to the Tx chain, the CPU goes to sleep if it finished the demodulation of the current symbol before
the end of the reception of the next one.

After the preamble detection, the receiver synchronizes to the transmitter using the algorithm described in~\cite{xhonneux2019low}.
The goal of the synchronization algorithm is to estimate and correct both the CFO and the STO using the preamble.
The fractional CFO is first estimated by multiplying the DFT $Y^{(2)}_k$ with the complex conjugate $\overline{Y}^{(1)}$ of the DFT of the previous upchirp~\cite{bernier2020low}:
    $\widehat{\lambda}_{\text{CFO}} = \frac{1}{2 \pi} \angleF \left( \sum^{2}_{i = -2} Y^{(2)}_{s + i} \cdot \overline{Y}^{(1)}_{s + i} \right)$,
where $s = \argmax_k \left| Y^{(2)}_k \right|$. Once the fractional CFO estimate $\widehat{\lambda}_{\text{CFO}}$ is obtained, the reference
signal $\overline{x}_0[n]$ is recomputed to directly include the correction term $e^{-j 2 \pi \frac{\widehat{\lambda}_{\text{CFO}} n}{N}}$.
The subsequent symbols are hence demodulated such that the fractional CFO is corrected. The receiver then iteratively estimates
the fractional STO using the following estimator from~\cite{xhonneux2019low}:
\begin{equation} \label{eq:estimator-lambda}
    \widehat{\lambda}_{\text{STO}} = -\real \left\lbrack \dfrac{e^{-j 2 \pi \tfrac{\widehat{L}_{\text{STO}}}{N}} Y_{i+1} - e^{j 2 \pi \tfrac{\widehat{L}_{\text{STO}}}{N}} Y_{i-1}}
                        {2 Y_i - e^{-j 2 \pi \tfrac{\widehat{L}_{\text{STO}}}{N}} Y_{i+1} - e^{j 2 \pi \tfrac{\widehat{L}_{\text{STO}}}{N}} Y_{i-1}} \right\rbrack .
\end{equation}
A first estimate $\tilde{\lambda}_{\text{STO}}$ is obtained using Eq.~\ref{eq:estimator-lambda} with the DFT $Y^{(4)}_k$ of the fourth upchirp of the preamble
and the approximation $\widehat{L}_{\text{STO}} \approx \argmax_k \left| Y^{(4)}_k \right|$, since the integer STO cannot be estimated yet.
The fractional STO, i.e., the intra-sample time offset, is corrected by updating the decimation index in the DFE among the $R$ available polyphases.
The receiver selects the decimation index that is the closest to the value $R \tilde{\lambda}_{\text{STO}}$.
With both fractional offsets corrected, the remainder of the preamble is demodulated, which allows to retrieve the integer CFO and STO.
Once $L_{\text{STO}}$ is estimated, the algorithm recomputes a more accurate estimate $\widehat{\lambda}_{\text{STO}}$ with $\widehat{L}_{\text{STO}}$ using Eq.~\ref{eq:estimator-lambda}
and updates the decimation index in the DFE.

After the processing of the preamble, the receiver starts the demodulation of the header and the payload.
The receiver however needs to be aligned with the boundaries of the first header symbol by correcting the integer STO.
To perform this correction in a simple manner, we treat the two adjacent ping-pong buffers as a single circular buffer of two symbols, as
illustrated in Fig.~\ref{fig:sw-memory}. The DSP routines of the CMSIS DSP library required by the demodulation (complex multiplication, FFT and magnitude computation) have been rewritten to properly handle this circular buffer.
The processor can hence start the demodulation of a symbol anywhere in the circular buffer, depending on $\widehat{L}_{\text{STO}}$.
The integer CFO is corrected by subtracting the estimate $\widehat{L}_{\text{CFO}}$ from the values of the demodulated symbols before the Gray demapping.

After the reception of the entire packet, the received bits are deinterleaved and decoded using a hard-decision Hamming decoder.
If the header indicates the presence of a CRC, the CRC value is computed on the payload and checked against the one present at the end of the packet.
If no errors are found, the payload bits are finally dewhitened and returned.

We finally note that due to the limited die size of the chip, our MCU design only embeds 32~kB memories. As a consequence,
we are unable to run the Rx chain for SFs higher than 9 as the total size of the both the program and FFT tables exceeds the available space in the PMEM.
Commercial MCUs however usually embed much larger memories (up to 512~kB).

\subsection{Evaluation of the Minimum CPU Frequency}

\begin{figure}[t]
    \begin{tikzpicture}[
    every axis/.style={ 
      ybar stacked,
      ymin=0,ymax=140000,
      bar width=24pt,
      xmin=0.5, xmax=3.5,
      xtick = {1,2,3},
      xticklabels={7, 8, 9},
      xticklabel style={yshift=-7pt},
      xlabel={Spreading Factor},
      xlabel style={yshift=-4pt},
      ylabel={\# CPU cycles per symbol},
      legend columns=-1,
      legend style={at={(0.03,0.9)},anchor=west}
    },
  ]
  
  \begin{axis}[bar shift=-14pt,hide axis]
  \addplot[fill=CBlue] coordinates
  {(1, 3136) (2, 6272) (3, 12544)};
  \label{plot:cycles_mod}
  \end{axis}
  
  \begin{axis}[bar shift=14pt]

  \addlegendimage{/pgfplots/refstyle=plot:cycles_mod}
  \addlegendentry{Modulation}

  \addplot[fill=CYellow] coordinates
  {(1, 5440) (2, 10880) (3, 21760)};
  \addlegendentry{Dechirping}
  \addplot[fill=COrange] coordinates
  {(1, 13056) (2, 23168) (3, 61632)};
  \addlegendentry{FFT}
  \addplot[fill=CPurple] coordinates
  {(1, 2880) (2, 5760) (3, 11520)};
  \addlegendentry{Magn.}
  \end{axis}

  \node[text width=2.5cm] (f1) at (1.35,1.5) {\footnotesize Minimum CPU freq.\\ in Rx: $20.9$~MHz};
  \node[text width=1cm] (f2) at (3.4,2.9) {\footnotesize $19.4$~MHz};
  \node[text width=1cm] (f3) at (5.7,4.4) {\footnotesize $23.2$~MHz};

  \node[text width=0.8] (Tx1) at (1.35-0.85,-0.2) {\small \textit{\textcolor{darkgray}{Tx}}};
  \node[text width=0.8] (Tx2) at (3.4-0.65,-0.2) {\small \textit{\textcolor{darkgray}{Tx}}};
  \node[text width=0.8] (Tx3) at (5.7-0.65,-0.2) {\small \textit{\textcolor{darkgray}{Tx}}};

  \node[text width=0.8] (Rx1) at (1.35+0.15,-0.2) {\small \textit{\textcolor{darkgray}{Rx}}};
  \node[text width=0.8] (Rx2) at (3.4+0.4,-0.2) {\small \textit{\textcolor{darkgray}{Rx}}};
  \node[text width=0.8] (Rx3) at (5.7+0.35,-0.2) {\small \textit{\textcolor{darkgray}{Rx}}};

  \draw[->,>=stealth,bend right=-45, color=CRed] (f1) to node[midway,above, rotate=45] {\scriptsize $T_S \times 2$} (f2.west);
  \draw[->,>=stealth,bend right=-45, color=CRed] (f2) to node[midway,above, rotate=45] {\scriptsize $T_S \times 2$} (f3.west);

  \end{tikzpicture}
    \caption{Number of CPU cycles required to modulate or demodulate LoRa symbols for different SFs.
    The demodulation stage is broken down in three substages: dechirping, FFT and magnitude computation of each FFT bin.
    The minimum CPU frequency required in Rx is obtained for $B = 125$kHz.}
    \label{fig:cycles}
\end{figure}
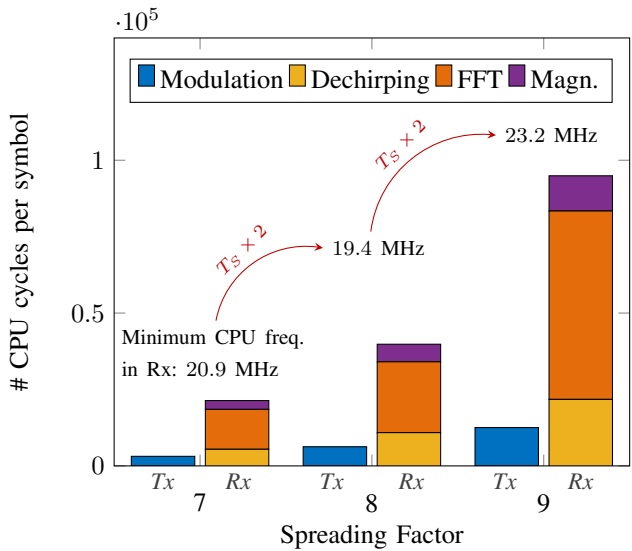

Finally, we analyze the efficiency of our implementation by determining the minimum CPU frequency needed to process LoRa packets in real-time.
Figure~\ref{fig:cycles} shows the number of CPU cycles required to modulate or demodulate a LoRa symbol for several spreading factors.
We observe that the demodulation of a symbol requires a greater number of cycles than the modulation. Whereas the modulation simply consists in
copying $N$ samples in memory, demodulating a symbol notably involves an FFT of complexity $\mathcal{O}(N \log{} N)$.
Consequently, the demodulation of symbols with higher SFs requires
more cycles than for lower SFs. However, the CPU has also more time to perform the FFT as the symbol duration $T_S$ depends on the SF.
Hence, the minimum CPU frequency for real-time operation does not vary much with the SF.
For $B = 125$~kHz, which is the typical bandwidth used in Europe~\cite{haxhibeqiri2018survey}, the minimum required frequency
increases from $20.9$~MHz to $23.2$~MHz when increasing the SF from 7 to 9.
Such frequencies are easily attained by ULP MCUs as they commonly have CPU frequencies in the range $32$ -- $180$~MHz~\cite{bol2021sleeprunner}.
We note that the minimum CPU frequency does not increase monotonically with the SF as the FFT implementation of the CMSIS library is more
efficient for FFT sizes with an even power of two.

\section{Experimental Performance Evaluation}
\label{sec:perfs}

\begin{figure}[t]
    \centering
    \includegraphics[width=0.95\linewidth]{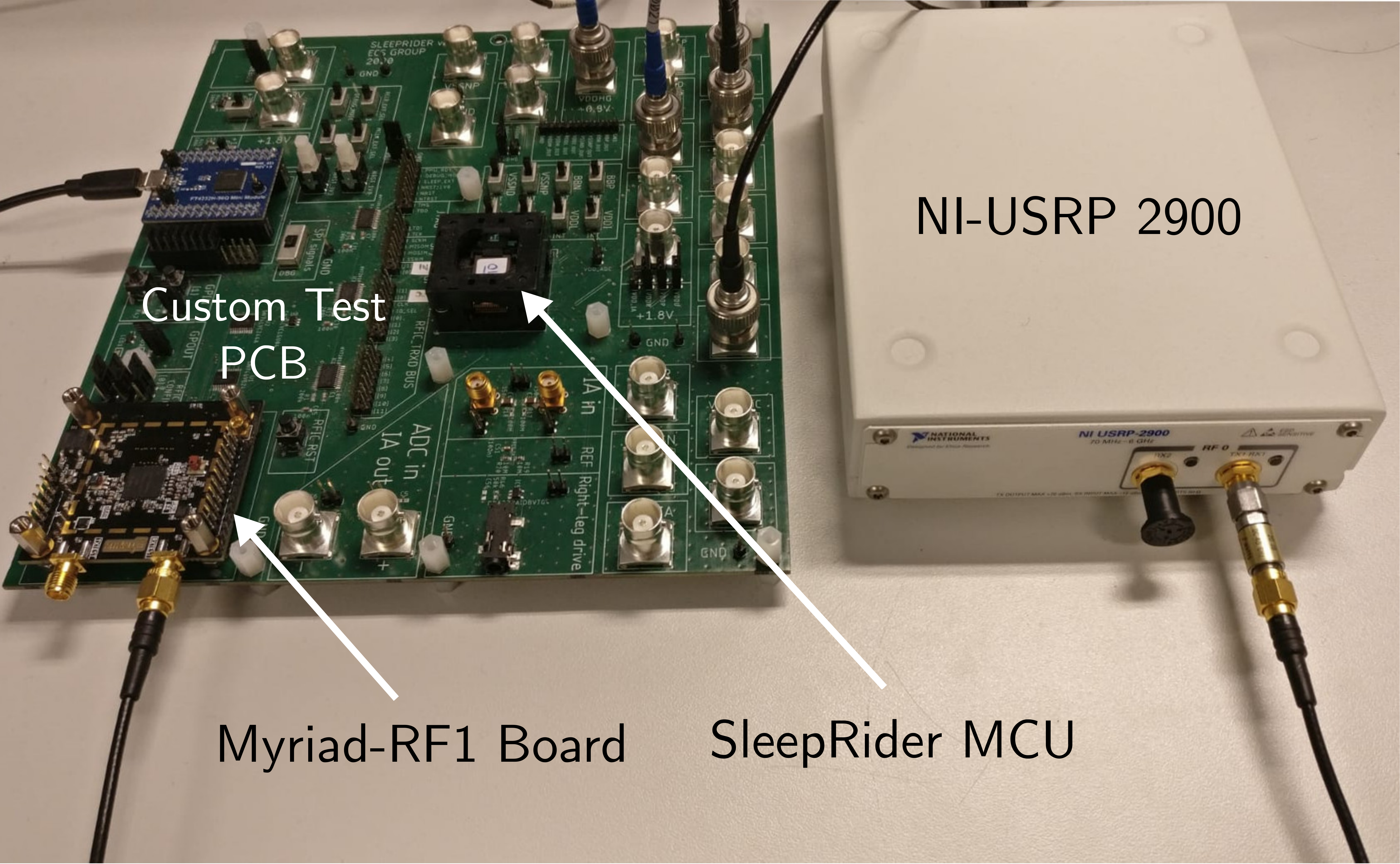}
    \caption{Testbed with the SleepRider MCU, the Myriad-RF1 reconfigurable transceiver and a NI-USRP 2900 running the LoRa GNU Radio SDR
    implementation of~\cite{tapparel2020open}.}
    \label{fig:testbed}
\end{figure}

We prototyped the architecture from Section \ref{sec:hw} within the SleepRider MCU in 28nm FDSOI~\cite{sleeprider2021}.
The fabricated MCU is integrated in a testbed to experimentally validate the proposed LoRa SDR LoRa and evaluate its performance.
The testbed is shown in Fig.~\ref{fig:testbed}. Beside the MCU, it also contains a LimeSDR Myriad-RF1 LMS6002D reconfigurable transceiver board
and a NI-USRP 2900.
The Myriad-RF1 is used as RF front-end and transfers complex baseband samples from or to the MCU on a 12-bit bus with a sampling clock driven by the DFE~\cite{lms6002d}.
The USRP is driven by a computer running the LoRa GNU Radio SDR implementation of~\cite{tapparel2020open}, which is compatible with commercial LoRa radios.

We validated the functionality of both the Tx and Rx chains in the testbed for the SFs $7$ and $8$, $B = 125$~kHz and $\text{CR} = \nicefrac{4}{8}$,
which gives raw data rates of 3.4 and 1.9~kbps, respectively.
The frequency of the CPU is set at $64$~MHz (generated internally) and the frequency of the external DFE clock is $2$~MHz.
In Tx, this clock is divided by 16 to obtain an internal DFE clock at $125$~kHz. In Rx, the internal DFE clock is kept at $2$~MHz and the decimation stage
downsamples the received signal after low-pass filtering by $R = 16$.

\renewcommand{\arraystretch}{1.15}
\begin{table}[t]
    \centering
    \begin{tabular}{|c|c|c|c|}
        \hline
        \textbf{Stage} & \textbf{\% of time} & \textbf{Power {\footnotesize (mW)}} & \textbf{Energy {\footnotesize (mJ)}}\\
        \hline
        Demodulation & 30.32\% & 0.87 & 27.19 (62.69\%) \\
        \hline
        Synchronization & 0.44\% & 0.87 & 0.39 (0.91\%) \\
        \hline
        Decoding & 0.17\% & 0.73 & 0.13 (0.30\%) \\
        \hline
        Wait for Interrupt & 69.07\% & 0.22 & 15.67 (36.11\%) \\
        \hline
        \textit{Average Total} & 100\% & 0.42 & 43.38 (100\%) \\
        \hline
    \end{tabular}
    \caption{Power characterization of the Rx chain when receiving a LoRa packet with a 11-byte payload
    for $B = 125$~kHz, $\mathrm{SF} = 8$, $\text{CR} = \nicefrac{4}{8}$ and a CPU frequency of $64$~MHz.}
    \label{tab:perfs}
\end{table}

We first analyze the energy consumption of the MCU when it receives a LoRa packet from the USRP
for $\textrm{SF} = 8$ and a payload length of 11 bytes. The RF front-end is excluded from the power measurements.
Table~\ref{tab:perfs} gives the proportion of time spent and the energy consumed by the MCU in each stage of the Rx chain, including the state when the
CPU has finished demodulating the current symbol and waits for an interrupt of the DFE to process the next one.
In this breakdown, we include the demodulation of the preamble symbols in the demodulation stage. The synchronization stage thus only contains the additional computations needed for the estimation and correction of the STO and CFO.
Overall, the Rx chain consumes an average power of $0.42$~mW with a processing energy of $493$~µJ/bit.

\begin{figure}[t]
    \pgfplotsset{
    legend image with text/.style={
        legend image code/.code={%
            \node[anchor=center] at (0.3cm,0cm) {#1};
        }
    },
}

\begin{tikzpicture}

	\pgfplotsset{grid style={dashed}}

    \begin{semilogyaxis}[
        xlabel = {SNR (dB)},
        ylabel = {Packet Error Rate},
        ylabel near ticks,
        xlabel near ticks,
        xmin = -16, xmax = -6,
        ymin = 1e-2, ymax = 1,
        grid = both,
        width=\linewidth,
        height = 7cm,
        legend columns=2,
        transpose legend,
        legend style={at={(0.04,-0.28)},anchor=west},
        legend style={nodes={scale=0.65, transform shape}}
        ]

        \addlegendimage{legend image with text=\textbf{$\mathrm{SF} = 7$:}}
        \addlegendentry{}

        \addlegendimage{legend image with text=\textbf{$\mathrm{SF} = 8$:}}
        \addlegendentry{}

        \addplot[CBlue, very thick, dotted, mark=none, mark options={scale=1, solid}] table [x=SNR, y=SER, col sep=comma] {th_per_sf7.csv};
        \addlegendentry{Theoretical}

        \addplot[COrange, very thick, dotted, mark=none, mark options={scale=1, solid}] table [x=SNR, y=SER, col sep=comma] {th_per_sf8.csv};
        \addlegendentry{Theoretical}

        \addplot[CBlue, thick, dashed, mark=square, mark options={solid, scale=1}] table [x=SNR, y=SER, col sep=comma] {sim_per_sf7.csv};
        \addlegendentry{Simulation}

        \addplot[COrange, thick, dashed, mark=square, mark options={solid, scale=1}] table [x=SNR, y=SER, col sep=comma] {sim_per_sf8.csv};
        \addlegendentry{Simulation}

        \addplot[CBlue, thick, solid, mark=o, mark options={scale=1}] table [x=SNR, y=SER, col sep=comma] {exp_per_sf7.csv};
        \addlegendentry{Experimental}

        \addplot[COrange, thick, solid, mark=o, mark options={scale=1}] table [x=SNR, y=SER, col sep=comma] {Texp_per_sf8.csv};
        \addlegendentry{Experimental}
	\end{semilogyaxis}

\end{tikzpicture}%
    \caption{Theoretical, simulation and experimental PERs when receiving LoRa packets with a 11-byte payload
    for different spreading factors, $B = 125$~kHz, and $\text{CR} = \nicefrac{4}{8}$.}
    \label{fig:per}
\end{figure}
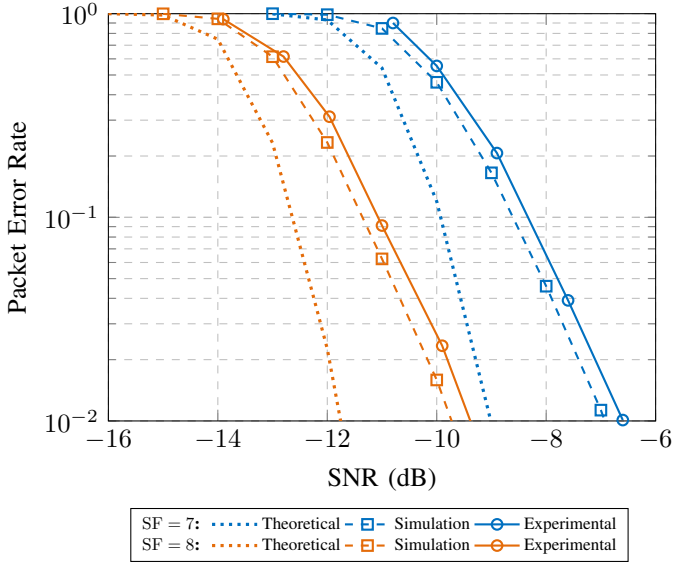

Finally, we assess the performance of our SDR Rx chain by measuring in the testbed the packet error rate (PER) versus the signal-to-noise ratio (SNR).
In this experiment, we generate different SNR values by sweeping the Tx gain of the USRP. 1000 LoRa packets are transmitted per SNR level.
The SNR is measured at the receiver side, similarly to~\cite{tapparel2020open}.
Figure~\ref{fig:per} shows both the simulation and experimental PERs as well as the theoretical PER of a perfectly synchronized receiver.
The simulation results are obtained in MATLAB by simulating the Rx chain (synchronization, demodulation
and decoding) in floating point with realistic CFO and STO values.
For a target PER of $10^{-2}$ and both $\mathrm{SF} = 7$ and $\mathrm{SF} = 8$, we observe that our SDR implementation is only $0.2$~dB away from the simulation results,
and requires only $2.2$~dB higher SNR than an ideal receiver.
Moreover, the receiver fully benefits from the $2.8$~dB spreading gain when increasing the SF from 7 to 8, in accordance with the theory.

\section{Conclusion}

Compared to conventional radios implemented in hardware,
IoT SDRs are a promising solution to extend the lifetime of IoT sensors.
In this work, we demonstrate a ULP MCU embedding a reconfigurable radio digital front-end and a Cortex-M4 CPU that is capable of transmitting
and receiving LoRa packets in real-time. Our software implementation of the LoRa PHY only uses the generic SIMD DSP instructions of the CPU
to modulate and demodulate symbols.
For a symbol bandwidth $B = 125$~kHz and a spreading factor $\mathrm{SF} = 8$, the Rx chain requires a minimum CPU frequency of $20$~MHz
and consumes only $0.42$~mW on average.
To the best of our knowledge, this work is the first to experimentally demonstrate a complete and functional sub-mW SDR implementation of a modern LPWAN protocol on a ULP MCU.
In future work, we will also implement the SigFox standard.

\bibliographystyle{IEEEtran}
\bibliography{IEEEabrv,paper}

\end{document}